\documentclass[3p,times,twocolumn]{elsarticle}

\usepackage{ecrc}

\volume{00}

\firstpage{1}

\journalname{Nuclear and Particle Physics Proceedings}

\runauth{S. Wykes et al.}

\jid{nppp}

\jnltitlelogo{Nuclear and Particle Physics Proceedings}

\usepackage{amssymb}

\usepackage{amsthm}
\usepackage{mathtools}
\usepackage{natbib}
\usepackage{graphicx}
\usepackage{graphics}

\usepackage[figuresright]{rotating}

\begin{document}

\begin{frontmatter}

\dochead{}

\title{UHECR propagation from Centaurus\,A}

\author[um]{Sarka Wykes\corref{cor1}}
\ead{sw@extragalactic.info}

\author[dias]{Andrew M. Taylor}

\author[jbca]{Justin D. Bray}

\author[herts]{Martin J. Hardcastle}

\author[leeds]{Michael Hillas}

\address[um]{Department of Physics and Astronomy, University of Manitoba, Winnipeg R3T 2N2, Canada}

\address[dias]{Dublin Institute for Advanced Studies, 31 Fitzwilliam Place, Dublin, Ireland}

\address[jbca]{JBCA, School of Physics and Astronomy, University of Manchester, Manchester M13 9PL, UK}

\address[herts]{School of Physics, Astronomy and Mathematics, University of Hertfordshire, College Lane, Hatfield, Hertfordshire AL10 9AB, UK}

\address[leeds]{School of Physics and Astronomy, University of Leeds, Leeds LS2 9JT, UK}

\cortext[cor1]{Corresponding author}

\begin{abstract}

In the light of the recently predicted isotopic composition of the kpc-scale jet in Centaurus\,A, we re-investigate whether this source could be responsible for some of the ultra-high energy cosmic rays detected by the Pierre Auger Observatory. We find that a nearby source like Centaurus\,A is well motivated by the composition and spectral shape, and that such sources should start to dominate the flux above $\sim4$\,EeV. The best-fitting isotopes from our modelling, with the maximum $^{56}$Fe energy fixed at $250$\,EeV, are of intermediate mass, $^{12}$C to $^{16}$O, while the best-fitting particle index is\,$2.3$. 
\end{abstract}


\end{frontmatter}

\section{Introduction} \label{sect:introduction}

Ultra-high energy cosmic rays (UHECRs) are the highest-energy particles found in nature. Good recent reviews are offered by \cite{WAT14} and in the publications from the \emph{Cosmic ray origin -- beyond the standard models} 2016 conference\footnote{\tt http://www.crbtsm.eu} (these proceedings). No conclusive signal revealing the origin(s) of UHECRs has yet emerged.

The interpretation of the Pierre Auger Observatory data \citep[e.g.][]{ABR11, ABR13} strongly hints at a varying cosmic-ray composition as a function of energy, and an Auger $X_{\rm max}$ width-based study now also shows that the composition tends towards intermediate masses \citep{AAB14} at the highest energies, at least for the part of the sky covered by the Auger Observatory. Most recently, based on a correlation between the depth of shower maximum and the signal in the water Cherenkov stations of air showers registered simultaneously by the fluorescence and the surface detectors of the Auger Observatory, i.e.\,via a method relatively {\it robust to uncertainties in the hadronic models}, \cite{AAB16} have shown that the observed correlation in the energy range $3.2$ to $10$\,EeV is inconsistent with expectations for pure primary CR composition and also with light composition consisting of proton and helium nuclei only. 

There is mounting evidence that the sources of (U)HECRs must be local, within $\sim30$\,Mpc \citep[e.g.][]{ACK16,GLO17}. A strong need for local sources of CRs around $1$\,EeV comes from gamma-ray fluxes \citep{LIU16}. For CRs above these energies, the {\it Fermi}-LAT extragalactic gamma-ray background limits \citep{ACK16} demonstrate that if the EeV cosmic rays are protons, their contribution to the extragalactic gamma-ray background is problematic. The general requirement for relatively flat particle spectra \citep[e.g.][]{ALO14, TAY14a, TAY14b} is also alleviated for nearby sources.

Photodisintegration of nuclei from nearby sources will not have a significant effect -- except for $^{4}$He and possibly for $^{14}$N, $^{16}$O and $^{20}$Ne which are more fragile \citep[e.g.][Section\,\ref{sect:disintegration}]{HOO07} -- on a propagated spectrum, supporting local objects as promising candidates. 

Of the local objects, `radio-loud' active galactic nuclei (AGN) have long been considered potential UHECR sources for their radio flux densities and dimensions. Looking at radio flux alone to say something about cosmic-ray power is probably not correct, because we also require physically large lobes in the model in which the UHECR are accelerated and confined there. \cite{HAR10} has investigated this issue, concluding that there would be only $\sim20$ objects within $100$\,Mpc distance (and thus a handful within $30$\,Mpc) capable of accelerating particles to the same energies as the radio galaxy Centaurus\,A.\footnote{Although the model discussed in that work was a proton-only one, the conclusions should still be valid. That would only increase Centaurus\,A's dominance of the sky since many of the faint sources in the so far best attempt at a homogeneous all-sky radio survey \citep{VEL12} should not be considered as possible UHECR sources.}

Hosted by the closest elliptical galaxy NGC\,5128, Centaurus\,A is the nearest ($3.8\pm0.1$\,Mpc; \citep{HARR10a}) radio galaxy; this proximity is particularly opportune in testing models of particle content, cosmic-ray acceleration and UHECR propagation from the source. \cite{WYK15a} have relied on realistic stellar populations of the parent elliptical in order to estimate isotopic abundances in Centaurus\,A's jet (see also Table\,\ref{tab:rates} and Section\,\ref{sect:entrainment}). Their derived total entrainment rate, by calculating nucleosynthetic yields of isotopes in stellar winds, of $\sim2.3\times10^{-3}$\,M$_{\odot}$\,yr$^{-1}$ suggests jet deceleration on kpc scales, and a low-density ($\sim1\times10^{-8}$\,cm$^{-3}$) particle content of the giant lobes of the radio galaxy. This material is solar-like, with protons,\footnote{Protons, while not a product of stellar nucleosynthesis, are the most abundant component of stellar winds by number, and in most cases also by mass.} $^{4}$He, $^{16}$O, $^{12}$C, $^{14}$N and $^{20}$Ne as the principal ingredients.

\cite{WYK13} have argued that most likely only nuclei above a charge threshold can be accelerated to $\ge$55\,EeV energies in Centaurus\,A. The maximum $^{56}$Fe energy achieved in this model with final energisation by stochastic processes in the large-scale lobes translates to a proton cutoff energy at the source of $\sim10$\,EeV. Here, we follow up on those studies, asking whether the input rate of the intermediate-mass nuclei can give the output in terms of the flux of UHECRs from Centaurus\,A and reproduce the spectrum measured by the Auger Observatory.

The plan of the paper is as follows. In Section\,\ref{sect:inputs}, we discuss the overall model and the relevant astrophysical parameters. In Section\,\ref{sect:flux+energetics}, we examine the energetics of the source and the fraction of the all-sky flux which Centaurus\,A might be contributing, and lay out the flux normalisation scheme. Section\,\ref{sect:spectrum+comp} is focused around the composition-dependent and composition-independent spectral fits. The key findings are summarised in Section\,\ref{sect:summ}.

Throughout the paper, we define the energy spectral indices $\alpha$ in the sense $S_{\!\nu}\propto\nu^{-\alpha}$ and particle indices $p$ as $n(E)\propto E^{-p}$.

\section{Model and astrophysical inputs} \label{sect:inputs}

We consider three stages through which particles may be produced and energised to UH energies in Centaurus\,A.

Stage 1: The jet-enclosed stars in Centaurus\,A release material, of a {\it range of species}.

Stage 2: Some fraction of this material is injected into the accelerator. This fraction is {\it species-dependent}.

Stage 3: The injected material is accelerated. This process is {\it rigidity-dependent}, not conditional upon species.

\subsection{Stage 1: Entrainment} \label{sect:entrainment}

We assume the cosmic-ray emission from Centaurus\,A to originate from material entrained into its jets and transported to its giant lobes, where it is available for further boosting to UH energies. From the quantity and isotopic composition of material released by stars enclosed within the northern jet of Centaurus\,A calculated by \citet{WYK15a}, we compute for each isotope the rate of particle entrainment for both lobes (Table\,\ref{tab:rates}). The $^{4}$He/$^{56}$Fe number rate ratio here is $\sim9190$, the $^{4}$He/$^{12}$C ratio is $\sim827$, and $^{4}$He/$^{16}$O is $325$.

\subsection{Stage 2: Injection} \label{sect:injection}

The injection process into the accelerator, and relative rates for different nuclear species, is a long-standing problem \citep[e.g.][]{MEY97, DRU00}. We propose a {\it phenomenological prescription} for obtaining a multi-species energy spectrum which scales the spectra in energy per nucleon by $Z^{2}/A$:

\begin{equation}
 \frac{E\,dN}{dE_{{\rm per}A}} = f_{\rm A} \frac{E\,dN}{dE}\,,
\end{equation}
where 
\begin{equation}
f_{\rm A} = f_{\rm SW}\,Z^{2}/A\,,
\end{equation}
with $f_{\rm SW}$ being the stellar wind abundance value, $A$ the atomic number, $E_{{\rm per}A}$ the energy per nucleon and $p$ the particle index. This is equivalent, for a power law, to scaling the spectra in energy per particle by $Z^2\,A^{(p-2)}$. For an index of $p=2.3$, this will change the $^{4}$He/$^{56}$Fe number rate ratio to $24.6$ after injection, and the $^{4}$He/$^{12}$C and $^{4}$He/$^{16}$O ratios to $2.2$ and $0.9$, respectively. The prescription defines the relative injection fraction of each species, but does not define the absolute fraction of the material that is accelerated. 
\begin{table}
 \centering
 \caption{For each abundant isotope, we give the cumulative mass entrained by Centaurus\,A's jet over the $560$\,Myr lifetime of its giant lobes (table $8$ in \citep{WYK15a}), the mean mass injection rate into both lobes over this period (assuming them to be identical), and the corresponding numerical rate $\dot{N}_{\rm A,ent}$.}
\begin{tabular}{lccc}
 \hline
 isotope & \multicolumn{1}{c}{mass entrained} & \multicolumn{1}{c}{mass rate} & \multicolumn{1}{c}{number rate} \\
  & \multicolumn{1}{c}{(single lobe)} & \multicolumn{1}{c}{(twin lobes)} & \multicolumn{1}{c}{(twin lobes)} \\
  & \multicolumn{1}{c}{($M_\odot$)} & \multicolumn{1}{c}{($M_\odot$\,yr$^{-1}$)} & \multicolumn{1}{c}{(s$^{-1}$)} \\
 \hline
 $^{1}$H & $7.4 \times 10^{4}$ & $2.6 \times 10^{-4}$ & $1.0 \times 10^{46}$ \\
 $^{3}$He & $3.1 \times 10^{1}$ & $1.1 \times 10^{-7}$ & $1.4 \times 10^{42}$ \\
 $^{4}$He & $2.7 \times 10^{4}$ & $9.6 \times 10^{-5}$ & $9.1 \times 10^{44}$ \\
 $^{12}$C & $9.5 \times 10^{1}$ & $3.4 \times 10^{-7}$ & $1.1 \times 10^{42}$ \\
 $^{14}$N & $7.2 \times 10^{1}$ & $2.6 \times 10^{-7}$ & $7.0 \times 10^{41}$ \\
 $^{16}$O & $3.3 \times 10^{2}$ & $1.2 \times 10^{-6}$ & $2.8 \times 10^{42}$ \\
 $^{20}$Ne & $5.5 \times 10^{1}$ & $2.0 \times 10^{-7}$ & $3.7 \times 10^{41}$ \\
 $^{22}$Ne & $5.0 \times 10^{0}$ & $1.8 \times 10^{-8}$ & $3.1 \times 10^{40}$ \\
 $^{24}$Mg & $1.8 \times 10^{1}$ & $6.4 \times 10^{-8}$ & $1.0 \times 10^{41}$ \\
 $^{26}$Mg & $2.7 \times 10^{0}$ & $9.6 \times 10^{-9}$ & $1.4 \times 10^{40}$ \\
 $^{28}$Si & $2.3 \times 10^{1}$ & $8.2 \times 10^{-8}$ & $1.1 \times 10^{41}$ \\
 $^{32}$S & $1.6 \times 10^{1}$ & $5.7 \times 10^{-8}$ & $6.8 \times 10^{40}$ \\
 $^{56}$Fe & $4.1 \times 10^{1}$ & $1.5 \times 10^{-7}$ & $9.9 \times 10^{40}$ \\
 \hline
\end{tabular}

 \label{tab:rates}
\end{table}

\subsection{Stage 3: Acceleration} \label{sect:acceleration}

Once the particles are injected into the giant lobes, at sufficient energy to be accelerated further, we assume that all remaining processes are solely rigidity-dependent. This ignores any further species-dependent collisional energy-loss processes, as the environment in the lobes is sufficiently sparse that such collisions should be rare. We assume that each species $A$ is accelerated to a power-law distribution
 \begin{equation}
  \frac{d\dot{N}_{\rm A,inj}}{dE} = f_{\rm A} \, E^{-p} e^{-E/E_{\rm max}}  \text{\,\,\,\,\,for $E > E_{\rm min}$}\,, \label{eqn:powerlaw}
 \end{equation}
where $\dot{N}_{\rm A,inj}$ is the rate at which particles of this species are injected into the acceleration mechanism, $f_{\rm A}$ is a normalisation constant, and the energy limits
 \begin{align*}
  E_{\rm max} &= E_{\rm Fe,max} \times Z / Z_{\rm Fe} \\
  E_{\rm min} &= E_{\rm H,min} \times Z / Z_{\rm H}
 \end{align*}
are purely rigidity-dependent. Based on the outcome from the stochastic acceleration model for the giant lobes by \cite{WYK13}, we adopt \mbox{$E_{\rm Fe,max} = 10^{20.4}$}\,eV ($250$\,EeV).

\subsubsection{Particle indices} \label{sect:indices}

Particle spectra from plausible acceleration scenarios in the jet -- magnetic reconnection (pc scales), diffusive shock acceleration (pc and kpc scales), shear acceleration (kpc scales) and stochastic acceleration (kpc scales) -- might lead to power-law spectra in the close proximity of the acceleration region with a particle index in the range $1.5-2.4$. The spectrum will steepen due to radiative losses as particles move away from the acceleration spot. 

The giant lobes can either show power-law spectra or, in special cases, peaked spectra, from stochastic acceleration. The peaked spectrum is as much a result of energy-dependent escape from the lobes as the energy-dependent acceleration rate; the peak represents the balance between the two rates. The injection of particles might occur at the centre of the accelerator for the peaked spectrum to be apposite, which is plausible when the jet is driving the turbulence (for Centaurus\,A, jet driving the turbulence has been considered by \cite{WYK13} and \cite{WYK14}).

\subsubsection{Neutrino and photon luminosities} \label{sect:neutrino}

The model of the source by \cite{WYK13} and \cite{WYK15a} does not lead to measurable ultra-high energy neutrino and photon fluxes: both the jet-stellar wind interaction regions in the jet as well as the turbulent environment of the giant lobes for the final acceleration to UH energies are media with too small a cross section for proton-proton or proton-photon collisions to be important, which means that a non-detection of ultra-high energy neutrinos and photons from the direction of Centaurus\,A does not rule out the radio galaxy as a source of UHECRs.

\subsection{Photodisintegration} \label{sect:disintegration}

To investigate the various decay channels of the photodisintegration, we convolved cross-sections from \cite{KHA05} with the CMB and EBL radiation fields, which gives the interaction length (i.e.\,energy-loss length). The EBL radiation field used is from \cite{FRA08}.
\begin{figure}
\includegraphics[angle=-90,width=1.05\linewidth]{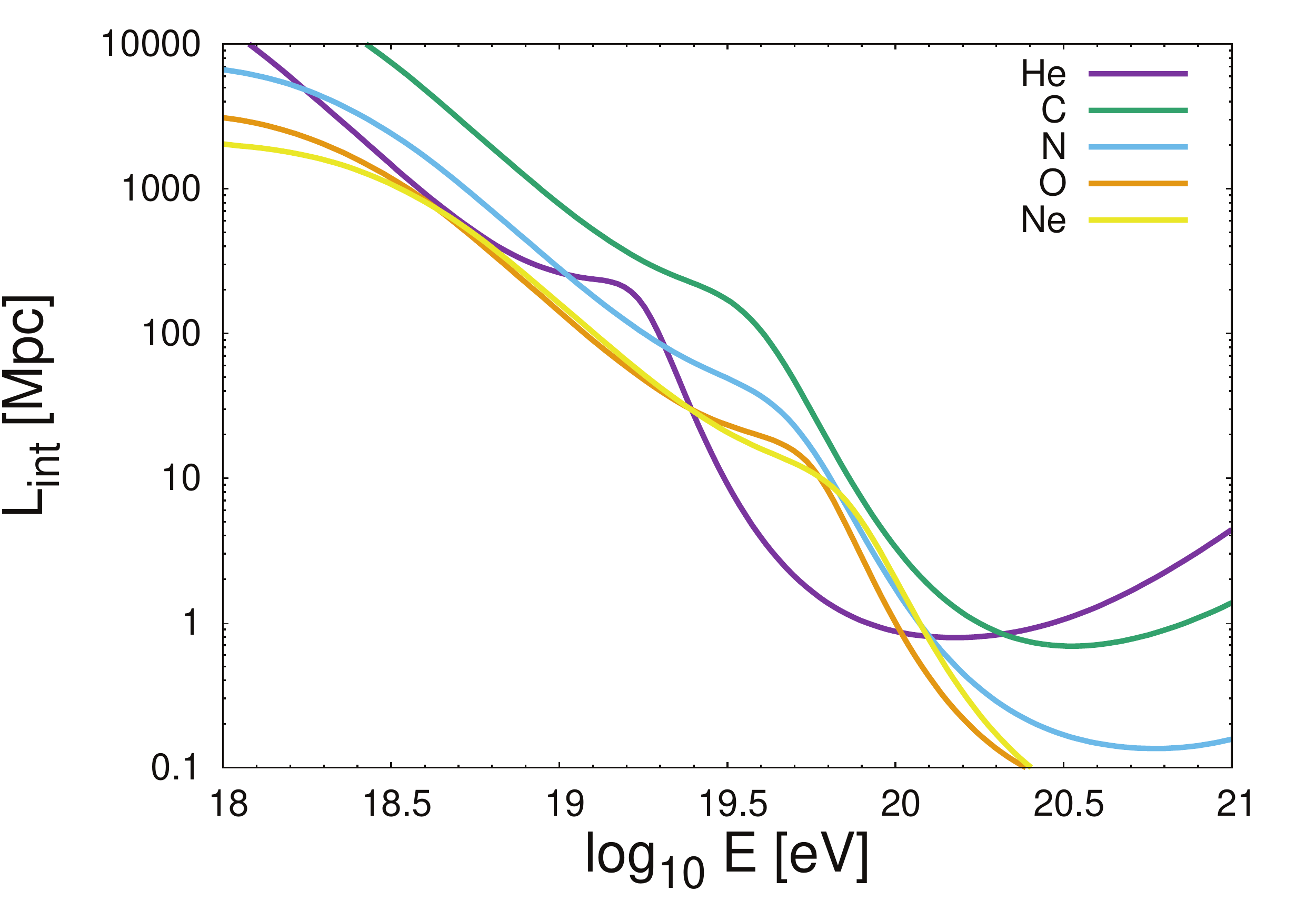}
\centering
\caption{Interaction length for five nuclei species (abundant at the source) due to photodisintegration on CMB and EBL, with the Khan photodisintegration cross section.} \label{fig:intlen_Khan}
\end{figure}

There are various exceptions to the general trend that nuclei are more robust at higher charge $Z$. Fig.\,\ref{fig:intlen_Khan} shows that of the intermediate-mass nuclei, $^{12}$C to $^{20}$Ne, $^{12}$C is relatively robust, with an interaction length of $\sim3.3$\,Mpc at 100\,EeV and $46$\,Mpc at 50\,EeV. $^{14}$N, $^{16}$O and $^{20}$Ne are more fragile at 100\,EeV; however, the robustness is higher at lower energies with, for example, the interaction length of $15.3$\,Mpc at $50$\,EeV for $^{16}$O and 12.6\,Mpc at 50\,EeV for $^{20}$Ne. The robustness of $^{14}$N is very comparable to $^{20}$Ne at $10^{19.7}$ to $10^{20.2}$\,eV; it is more robust than $^{20}$Ne outside these limits.

Given the proximity of our source, we focus on the initial steps of the disintegration. Our modelling shows that the second most important decay channel in the first photodisintegration step for $^{16}$O is feeding into $^{12}$C via $^{16}$O $\rightarrow$ $^{12}$C + $^{4}$He. For $^{20}$Ne, the most important (among many) decay channel in the first photodisintegration step leads again to stripping off an alpha particle, i.e. to a reaction $^{20}$Ne $\rightarrow$ $^{16}$O + $^{4}$He. The nitrogen isotope $^{14}$N feeds in the first step again into $^{12}$C, via $^{14}$N $\rightarrow$ $^{12}$C + 2$^{1}$H.
 
Thus, nuclei may reach the Earth largely unaffected. A marginally lighter arrival composition, enhanced in $^{12}$C and $^{4}$He levels (at the cost of $^{20}$Ne and $^{16}$O) is possible.

\section{Absolute flux and energetics} \label{sect:flux+energetics}

\subsection{Energetics} \label{sect:energetics}

To examine the overall energetics of cosmic-ray acceleration in Centaurus\,A, we first consider the case in which all material entrained in the jets is accelerated to high energies in its lobes, neglecting the species-dependent injection described in Section\,\ref{sect:injection}; i.e.\ assuming that \mbox{$\dot{N}_{\rm inj} = \dot{N}_{\rm ent}$}. Taking $\dot{N}_{\rm A,ent}$ for each species from Table\,\ref{tab:rates}, we can then calculate the normalisation $f_{\rm A}$ of its spectrum from equation\,\ref{eqn:powerlaw}. The particle index $p$ and the minimum energy $E_{\rm H,min}$ of the accelerated particles remain free variables, but we can constrain the latter, at least, to be greater than the mean thermal energy in the lobes at a temperature \mbox{$T \sim 2.0 \times 10^{12}$}\,K \citep{WYK13}, being  
 \begin{align}
  E_{\rm th}
   &= \frac{3}{2}\,k T \sim 0.26\,{\rm GeV} \label{eqn:Emin}
 \end{align}
where $k$ is the Boltzmann constant.

\begin{figure}
 \centering
 \includegraphics[width=1.01\linewidth]{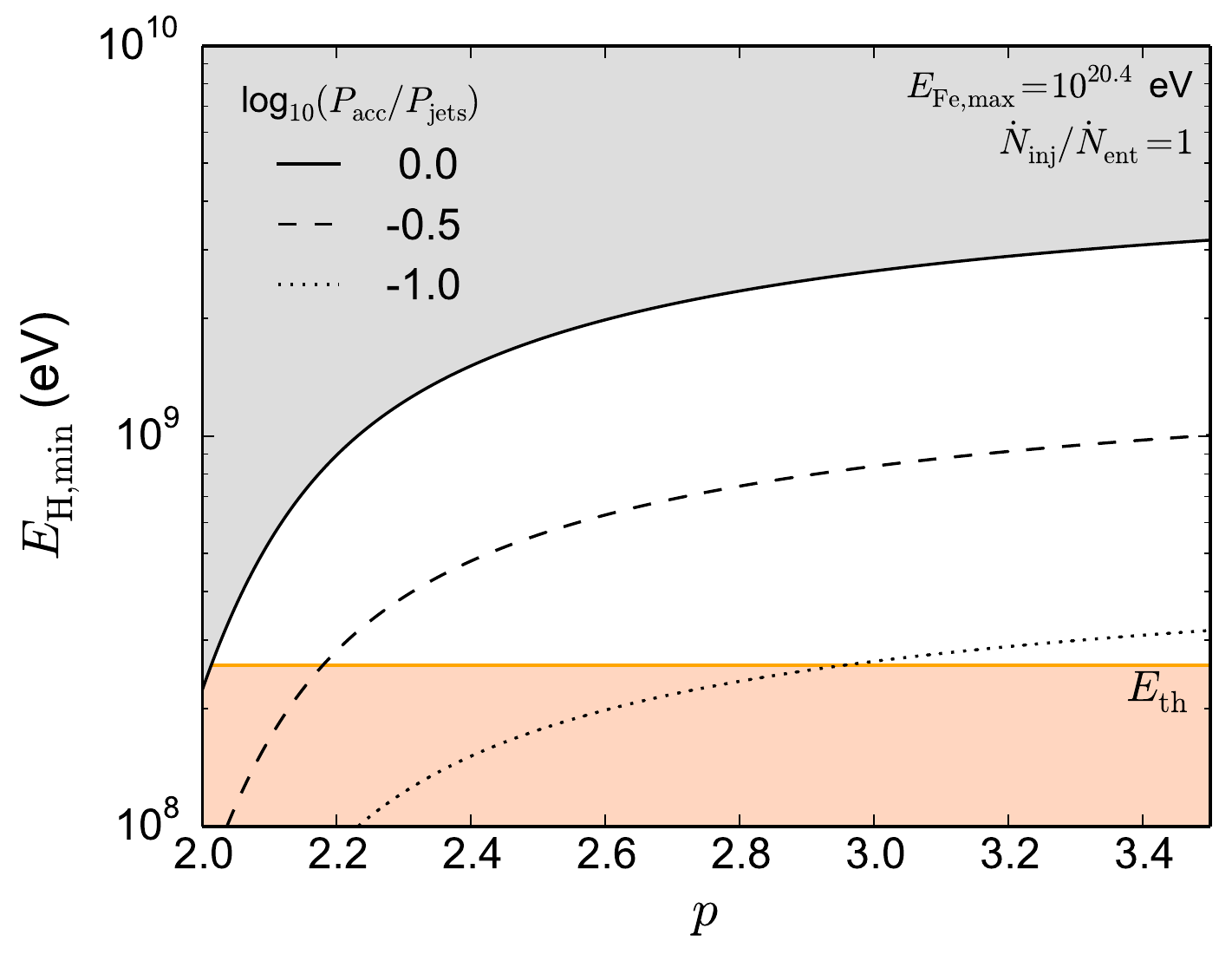}
 \caption{Fraction of jet power $P_{\rm jets}$ required for particle acceleration $P_{\rm acc}$, under the assumptions described in Section\,\ref{sect:energetics}, for different values of the particle index $p$ and the minimum energy $E_{\rm H,min}$ of the spectrum of accelerated particles. Lines follow parameter values with a constant acceleration efficiency $P_{\rm acc}/P_{\rm jets}$; the dotted line corresponds to the common assumption of $10\%$ acceleration efficiency. The shaded region at the top of the plot is excluded, as it requires more power for particle acceleration than is available from the jets. The horizontal line corresponds to the mean thermal energy $E_{\rm th}$ of particles in the lobes; the shaded region below this is excluded as it requires particles to be actively cooled below this energy.} 
 \label{fig:Pacc}
\end{figure}

From these parameters we can then obtain the total power used to accelerate particles of all species,
 \begin{equation}
  P_{\rm acc} = \sum_{\rm A} \int \! dE \, \frac{d\dot{N}_{\rm A}}{dE} \, E\,,
 \end{equation}
and compare it to the combined power supplied by both jets \mbox{$P_{\rm jets} = 10^{44}$}\,erg\,s$^{-1}$, based on the higher end of the historical jet power for a single jet of \mbox{$1-5 \times 10^{43}$}\,erg\,s$^{-1}$~\citep{WYK13, NEF15}. Fig.\,\ref{fig:Pacc} shows the ratio between these two values: the acceleration efficiency $P_{\rm acc}/P_{\rm jets}$. For a minimum energy close to $E_{\rm th}$, the particle index may be close to \mbox{$p = 2.0$} if the acceleration efficiency is close to $100\%$; however, if the acceleration efficiency is limited to \mbox{$\sim 10\%$}, the particle index must be \mbox{$\gtrsim 2.8$}.

Having calculated the normalisation of the accelerated particle spectrum in equation\,\ref{eqn:powerlaw}, and making the simplifying assumptions that cosmic rays are radiated isotropically, propagate ballistically, and do not interact while propagating, we can compute the resulting cosmic-ray flux at Earth as
 \begin{equation}
  \frac{d\dot{N}}{dE} = \sum_{\rm A} \frac{1}{4 \pi d^2} \frac{d\dot{N}_{\rm A}}{dE}\,, \label{eqn:ballistic_prop}
 \end{equation}
where $d$ is the distance to Centaurus\,A. This is a simplistic treatment, but for a close source, particle interactions are minimised (see also Section\,\ref{sect:disintegration}), and the most energetic particles experience relatively small deflections, so it gives an approximate absolute normalisation to the cosmic-ray flux from Centaurus\,A.  We have briefly examined the resulting fluxes and found that, for \mbox{$p \gtrsim 2.3$}, Centaurus\,A makes a negligible contribution to the all-sky UHECR flux, under the assumptions in this section.

\subsection{Cosmic-ray flux} \label{sect:crflux}

To investigate the conditions under which Centaurus\,A could contribute a significant fraction of the all-sky cosmic-ray flux, we next consider the case in which a small, species-dependent fraction of the material entrained in the jets is accelerated to high energies in the lobes, as described in Section\,\ref{sect:injection}. We fix the particle index to \mbox{$p = 2.63$}, matching the spectrum observed at energies above $10^{18.6}$\,eV ($4$\,EeV, traditionally called the `ankle'\footnote{We move away from this traditional nomenclature as it has no longer a sufficient physical basis.}) by the Auger Observatory\,\citep{VER16}, and determine the resulting cosmic-ray flux at Earth, assuming lossless, ballistic propagation per equation\,\ref{eqn:ballistic_prop}, as a fraction of the observed all-sky flux. For simplicity, we let $E_{\rm max} \rightarrow \infty$, which will have little effect on the energetics for this steep spectrum.

\begin{figure}
 \centering
  \includegraphics[width=1.03\linewidth]{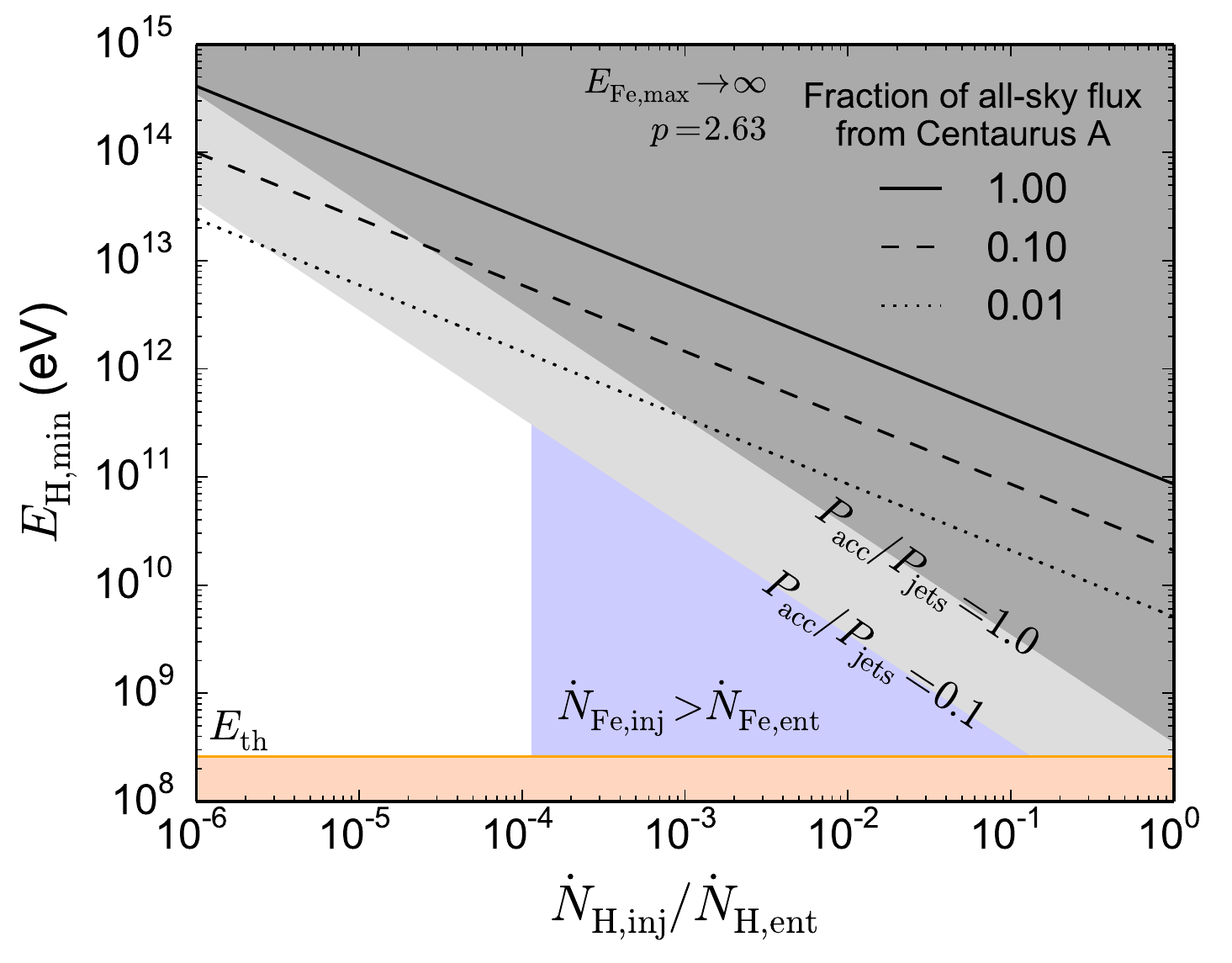}
 \caption{Fraction of the all-sky cosmic-ray flux beyond $4$\,EeV that would originate from Centaurus\,A, for the model laid out in Section\,\ref{sect:crflux}, for different values of the injection fraction $\dot{N}_{\rm inj}/\dot{N}_{\rm ent}$ and the minimum energy $E_{\rm min}$ of the spectrum of accelerated particles. Lines follow parameter values which result in a constant fraction of the all-sky cosmic-ray flux originating from Centaurus\,A, as described in the legend. The shaded region at the bottom of the plot is excluded, as in Fig.\,\ref{fig:Pacc}, because it requires particles to be cooled below the mean thermal energy in the lobes. The dark-shaded region in the upper right is closed out because it requires an acceleration efficiency \mbox{$P_{\rm acc}/P_{\rm jets} > 1$}; the light-shaded region corresponds to acceleration efficiencies exceeding 10\%.  Within the shaded region in the centre of the plot, it is not possible to strictly meet the prescription described in Section\,\ref{sect:injection}, as it leads to a disproportionately high iron content exceeding that available from entrained material.} 
 \label{fig:crfrac}
\end{figure}

Results are displayed in Fig.\,\ref{fig:crfrac}. For this particle index\,$p$, for Centaurus\,A to contribute significantly to the all-sky cosmic-ray flux, without exceeding the power available for acceleration from its jets, requires that only a small fraction \mbox{$\dot{N}_{\rm inj}/\dot{N}_{\rm ent} \lesssim 10^{-5}$} of the entrained particles are injected into the acceleration mechanism, and that they all be accelerated above a minimum energy \mbox{$E_{\rm min} \gtrsim 10^{13}$}\,eV. These limits may be relaxed if the acceleration mechanism results in spectral curvature, with a flatter spectrum at higher energies than at lower energies; or if non-rectilinear diffusion leads to a significant enhancement of the flux from this source.

A simple rectilinear flux normalisation as above seems challenged by the fact that any Centaurus\,A-related anisotropy of the Auger Observatory events is weak \citep[e.g.][]{AAB15}. A large fraction of the particles therefore may appear to be diffusing, which invariably also may alter the flux level (away from rectilinear). However, random deflections up to $\sim90^{\circ}$ would be sufficient to conceal a $\sim10\%$ contribution from Centaurus\,A to the all-sky flux, while only altering the flux level by a factor $\sim2$ (compared to rectilinear). Hence, it provides at a minimum an order-of-magnitude estimate for the normalisation.

\section{Propagated spectra and composition} \label{sect:spectrum+comp}

\subsection{Fitting methodology} \label{sect:fitting}

To determine the particle spectrum and compare this with the measurements of the Auger Observatory, we used a 3D Monte Carlo description of UHECR propagation as per \cite{TAY15}. Here, UHECR protons and nuclei are propagated through the cosmic microwave background (CMB) and cosmic infrared background (CIB) radiation fields, undergoing energy losses via photodisintegration, pair production, photo-pion collisions and losses due to cosmological redshift. In the present paper, we utilise the Khan photodisintegration cross section, based on phenomenological and microscopic models by \cite{KHA05}, and the description of the CIB spectral energy distribution by \cite{FRA08}. We implemented the hadronic models QGSJet II-4, EPOS-LHC and Sybill 2.1 into the analysis and fitting routines.

The assumed maximum $^{56}$Fe energy at the source \citep[][Section\,\ref{sect:acceleration}]{WYK13} of $250$\,EeV (not inconsistent, within the $1\sigma$ error margins, with the so far highest-energy event, observed by Fly's Eye, of $320\pm93$\,EeV; \citep{BIR95}) translates to a proton cutoff energy at the source of $9.6$\,EeV. Similar maximum proton energies at sources have been suggested by e.g.\,\cite{ALL08}, \cite{ALL09}, \cite{ALO11} and \cite{ALL12}. Since \citep{WYK15a} have predicted the abundances at source of all the elements and isotopes, and we adopt a prescription for them, there are no free parameters in the isotopic composition. 

As a first approach, we scanned over the range spanned by the hadronic models, with the minimum in the likelihood, within this range, being the value adopted, and over the particle spectral index. The results are presented in Section\,\ref{sect:unnormalised}. 

In the second approach, we left the normalisation free and the $\chi^2$ minimised for spectral data fit in the energy region $>10^{18.6}$\,eV ($4$\,EeV). No parameters were scanned over for this result which we discuss in Section\,\ref{sect:normalised}.

Note that while understanding the low-energy abundances at a given energy is paramount, we need to normalise to as high an energy as possible to minimise the complications arising from propagation through extragalactic and Galactic magnetic fields. The normalisation is therefore a best fit to the data above $10^{18.6}$\,eV.

\subsection{`Unnormalised', composition-dependent spectra} \label{sect:unnormalised}

\begin{figure}
\includegraphics[width=1.0\linewidth]{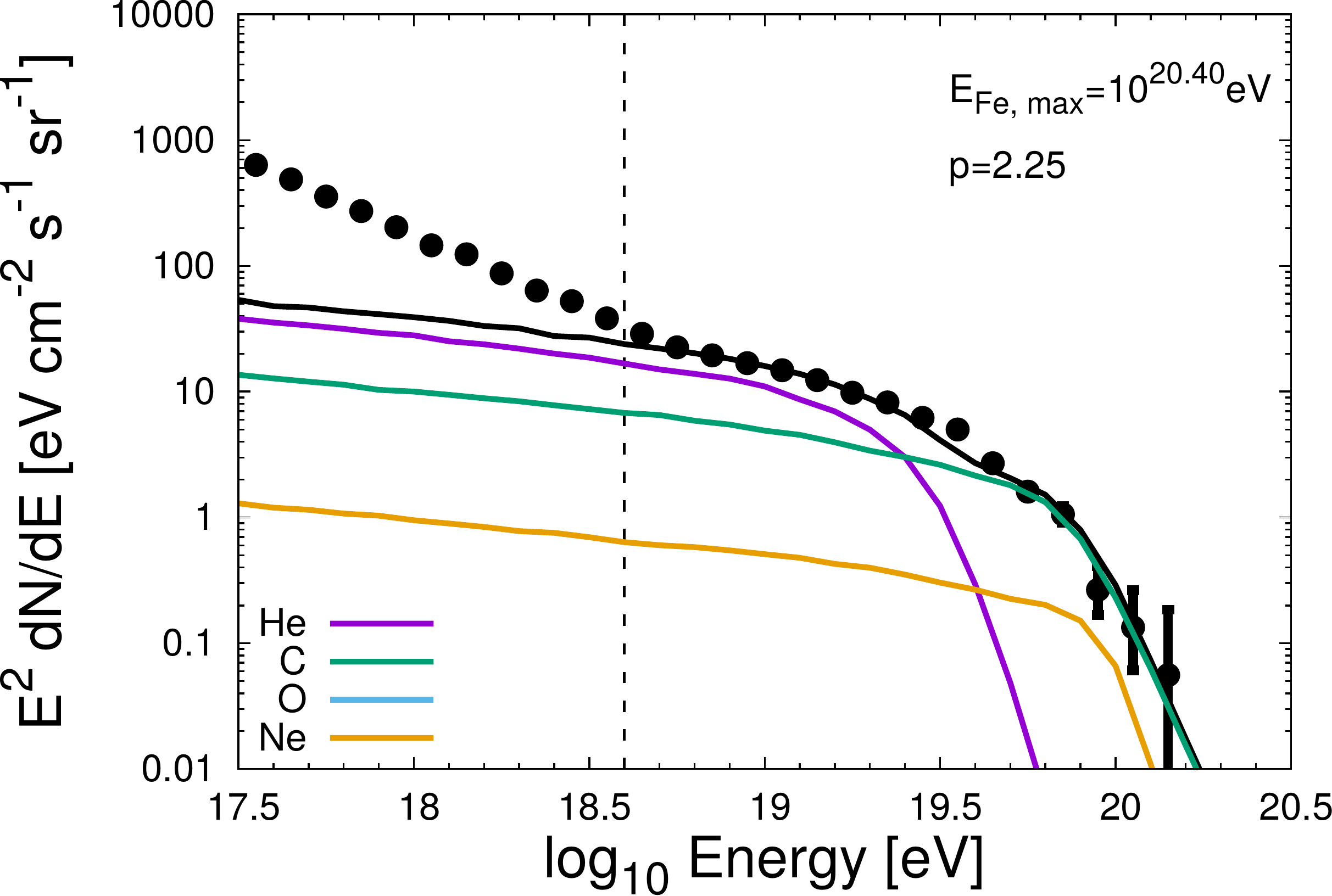}
\centering
\caption{Propagated spectra with $^{4}$He, $^{12}$C, $^{16}$O and $^{20}$Ne (solid lines), with the maximum energy at the source fixed at $^{56}$Fe$_{\rm max}=250$\,EeV, and with zero propagation magnetic field. Here, the overall flux is normalised to the Auger data (from \citep{VER16}). The errors on the data points shown are $1\sigma$ errors. The vertical dashed line indicates the point in the data where a spectral hardening occurs.} 
\label{fig:Spectrum_Auger_he4+c12+o16+ne20_Khan_Femax_fixed}
\end{figure}
Below, we describe the effect of normalising to the Auger data, i.e. a case without physically justified normalisation. Apart from the overall normalisation, also the composition ratios and the injection particle index were left free in the Monte Carlo scan. Each plot adopts a particular admixture abundant species set. For each admixture case considered, the global best-fit result is shown. We did not include $^{14}$N in the analysis as only $\sim7$ species can in principle be separated out by the current hadronic models, and to extract most of the data it is favourable to use logarithmically evenly-spread species.

\begin{figure}
\includegraphics[width=1.0\linewidth]{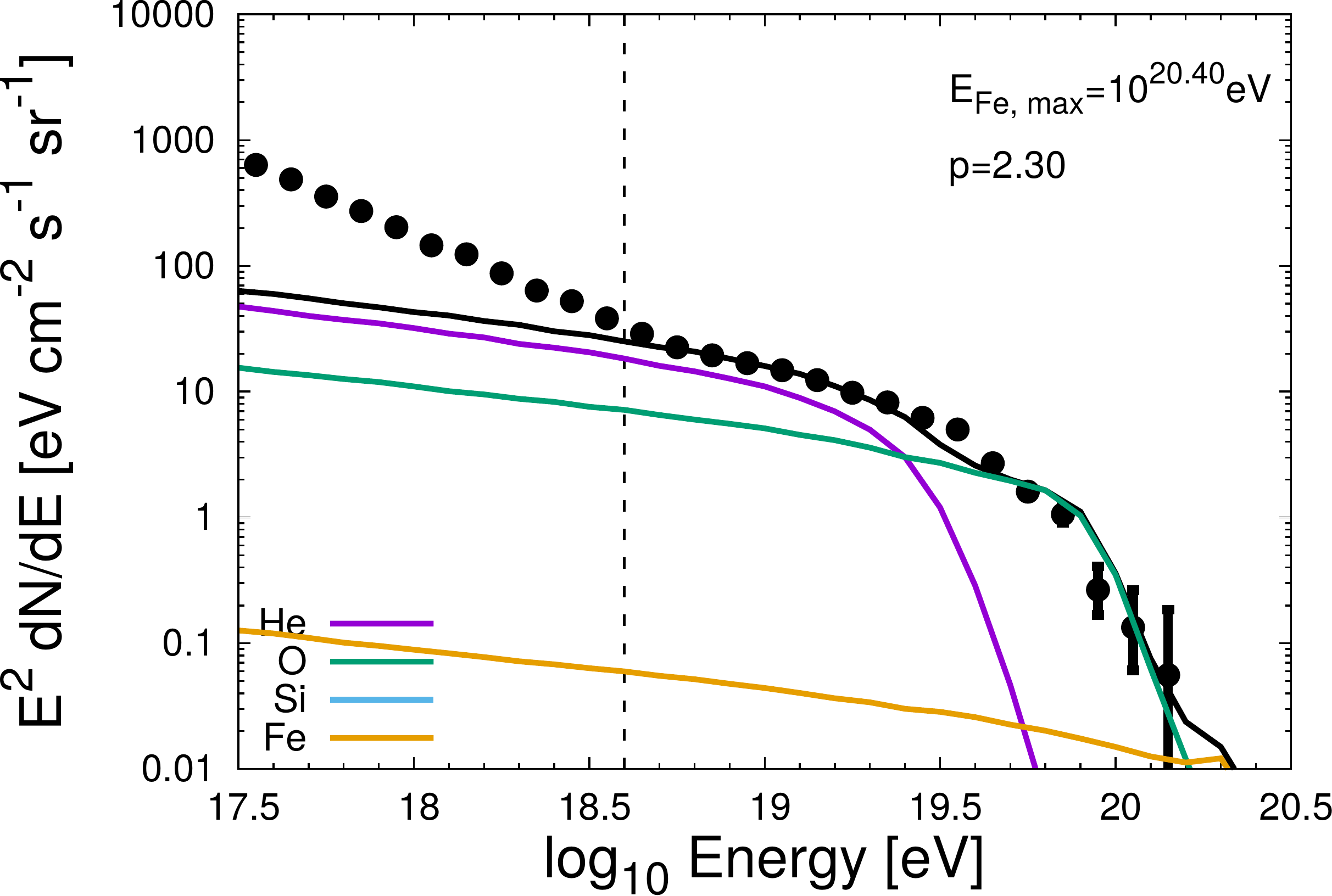}
\centering
\caption{As in Fig.\,\ref{fig:Spectrum_Auger_he4+c12+o16+ne20_Khan_Femax_fixed}, but for propagated spectra with $^{4}$He, $^{16}$O, $^{28}$Si and $^{56}$Fe.} \label{fig:Spectrum_Auger_he4+o16+si28+fe56_Khan_Femax_fixed}
\end{figure}
\begin{figure}
\includegraphics[width=1.0\linewidth]{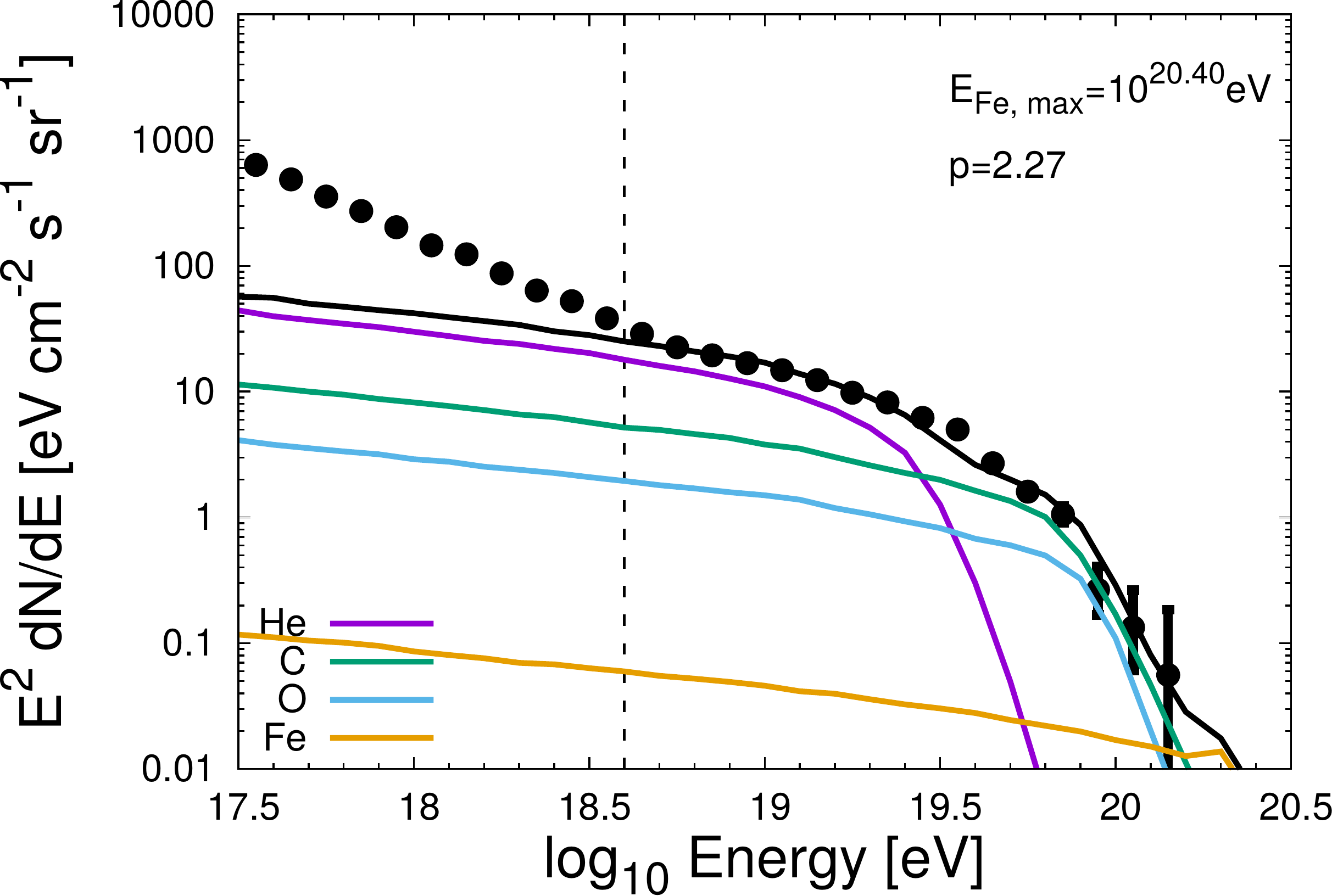}
\centering
\caption{As in Fig.\,\ref{fig:Spectrum_Auger_he4+c12+o16+ne20_Khan_Femax_fixed}, but for propagated spectra with $^{4}$He, $^{12}$C, $^{16}$O and $^{56}$Fe.} \label{fig:Spectrum_Auger_he4+c12+o16+fe56_Khan_Femax_fixed}
\end{figure} 

Figs\,$4-6$ show that, from the plotted species, $^{12}$C and/or $^{16}$O is required to match the Auger data at energies $\sim10^{19.5}$\,eV ($32$\,EeV). It is not possible to make a strong statement about $^{12}$C and $^{16}$O in particular: the species are too closely spaced in mass number to distinguish between them. The $^{12}$C spectrum has a break slightly `earlier' than the $^{16}$O spectrum, making a larger fraction of $^{12}$C than $^{16}$O preferable in the admixture in order to find agreement with the downturn feature from $10^{19.85}$ and $10^{19.95}$\,eV. The earlier downturn of $^{12}$C than $^{16}$O is a Lorentz factor effect here; it is not due to, for example, the $^{16}$O $\rightarrow$ $^{12}$C + $^{4}$He photodisintegration (which could potentially lead to additional $^{12}$C fluxes, see Section\,\ref{sect:disintegration}). Essentially, for a given energy, $^{12}$C has a slightly larger Lorentz factor than $^{16}$O, so can interact with somewhat lower-energy photons. The downturn is due to onset when interactions with CMB + CIB photons become possible.

The vertical dashed line in Figs\,$4-6$ at $10^{18.6}$\,eV is overlaid to stress that the data itself shows a new (harder) component that starts at these energies. Our source results begin to dominate there since our fluxes are power laws, and the lack of break in our (ballistic) results naturally has the effect that the power-law dominates at energies above the dashed line.

\subsection{Normalised, composition-independent spectra} \label{sect:normalised}

The results from Section\,\ref{sect:unnormalised}, in which the composition ratios are left free to float and where the intermediate to heavy ratios appear favoured, together with the solar low-energy cosmic-ray composition data, suggest a selective injection process.

\begin{figure}
\includegraphics[width=1.0\linewidth]{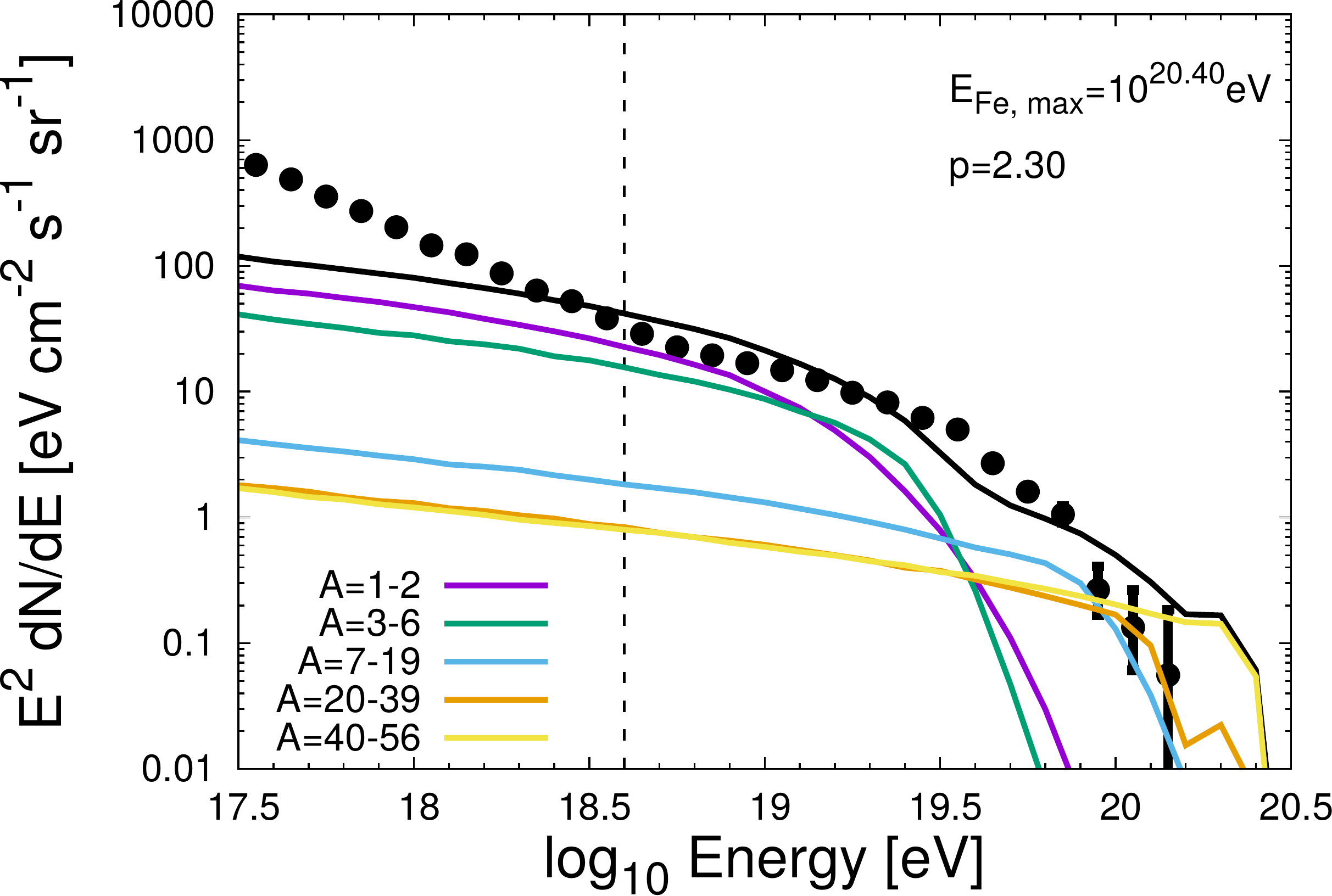}
\centering
\caption{Propagated spectra with $^{4}$He, $^{16}$O, $^{28}$Si and $^{56}$Fe (solid lines), with the maximum energy at the source fixed at $^{56}$Fe$_{\rm max}=250$\,EeV and the spectral index fixed at $2.3$, and with zero propagation magnetic field. Here, the overall flux is normalised based on the injection prescription (Section\,\ref{sect:injection}). The errors on the data points shown are $1\sigma$ errors. The vertical dashed line indicates the point in the data where a spectral hardening occurs. The kink at high energies is an artefact of the Monte Carlo method employed; due to the low statistics, the results are liable to Poisson noise.} \label{fig:Spectrum_Auger_normalised}
\end{figure}

Fig.\,\ref{fig:Spectrum_Auger_normalised} depicts the spectral outcome when Centaurus\,A is adopted as the dominant source of UHECRs above $10^{18.6}$\,eV ($4$\,EeV), for species in the ratios obtained by our injection prescription (Section\,\ref{sect:injection}). The slight departures from the Auger data demonstrate the need for other species, in the intermediate to heavy range, to be considered as well. The entrained composition, after scaling per our injection prescription (Section\,\ref{sect:injection}) does not contain relatively enough intermediate-mass nuclei to fit the observed spectrum. Protons and $^{4}$He exceed the spectrum at low energies, and $^{56}$Fe exceeds the spectrum at high energies, and the CNO in between is not as abundant as it needs to be.

The inclusion of plausible intergalactic magnetic fields of $\sim1$\,nG strength and with $0.1$\,Mpc coherence length may lead to an apparently increased surface brightness of the source at low energies, but is not expected to affect the fits at the highest energies in a strong way.

Centaurus\,A as discussed throughout this work could be representing nearby UHECR sources, in which case the total CR luminosity would be shared out amongst the different sources. This would also help alleviate the CR anisotropy concerns.

\section{Summary} \label{sect:summ}

The main results of this paper are as follows.

(1) Centaurus\,A and other nearby objects are well motivated as a source of UHECRs by the composition and spectral shape and start to dominate the CR flux above $\sim4$\,EeV. The best-fitting isotopes from our modelling, with the maximum $^{56}$Fe energy at the source fixed at $250$\,EeV, are of intermediate mass, $^{12}$C to $^{16}$O, although we cannot make a strong statement on $^{12}$C versus $^{16}$O (or $^{14}$N) due to close spacing in mass number. 

(2) Photodisintegration of nuclei is largely unimportant for a quasi-rectilinear particle transport from the source, except for a modest disintegration of $^{14}$N, $^{16}$O and $^{20}$Ne which will marginally enhance $^{12}$C and $^{4}$He levels at lower energies.

(3) The best-fitting power-law particle spectral indices, from an approach which considers composition-dependent spectra and artificially normalises to the Auger data, cluster around $2.3$, compatible with plausible acceleration scenarios at the source. The quantity of material accelerated to the highest energies in Centaurus\,A must be less than the material entrained from jet-enclosed stars, otherwise the particle spectral index is required to be too steep.

(4) Composition-independent spectra, with normalisation relying on our phenomenological prescription for injection, demand that additional isotopes, in the intermediate to heavy range, be considered.

In the next paper, we will also consider a range of extragalactic magnetic fields and the effects of Galactic fields on the propagation.

\section*{Acknowledgements}
We thank T. Jones, L. Drury, D. Ryu, P. Blasi, C. O'Dea, D. Caprioli and R. Gleisinger for helpful discussions. AMT acknowledges a Schr\"odinger Fellowship at DIAS. JDB acknowledges support from ERC-StG 307215 (LODESTONE).

\end{document}